\documentclass{evn2002}
\begin{document}
 \def\dpon{\hbox{$.\!\!^\circ$}}
   \title{ VLBI observations in Cluster-Cluster mode at 1.6 GHz}

   \author{M. J. Rioja\inst{1} \and  
           R. W. Porcas\inst{2} \and J.-F. Desmurs\inst{1} \and
           W. Alef\inst{2} \and L.I. Gurvits\inst{3} \and 
           R.T. Schilizzi\inst{3}
          }

   \institute{Observatorio Astron\'omico Nacional, Apartado 1143, E-28800 Alcal\'a de Henares, Madrid, Spain
         \and
             Max-Planck-Institut f\"ur Radioastronomie, Auf dem H\"ugel 69, D-53121 Bonn, Germany
         \and
             Joint Institute for VLBI in Europe, Postbus 2, 7990 AA Dwingeloo, The Netherlands
             }
   \abstract{
We describe, and present preliminary results from the analysis of 1.6\,GHz VLBI observations, made in 
Cluster-Cluster mode between subgroups of antennas from the WSRT, 
VLA and MERLIN arrays.
We observed for two 6-hour periods on consecutive days, 
simultaneously monitoring 4 compact radio sources 
with angular separations ranging from $\sim$1 to 9 degrees.
The data were recorded in a non-standard MkIII mode using special set-ups, and
correlated at the MPIfR MkIIIa correlator.
The preliminary results from standard and phase reference analyses
are very encouraging. The experiment has provided a useful database for investigating 
the potential of this new observing technique at low frequencies.
   }
 \authorrunning{M.J. Rioja et al.}
   \maketitle
\section{Introduction}

The Cluster-Cluster (Cl-Cl) or multiview VLBI technique is a
development of the observational capabilities of VLBI,
where single telescopes are replaced by sites with multiple
antenna elements (a ``cluster'') fed from a common local oscillator. The
Cl-Cl observing mode allows {\bf simultaneous} observations
of multiple sources on the separate ``sub-baselines'' between
sub-elements from one site to another, irrespective of their angular
separation. 

Hemenway (1974) describes such observations between 2 sites, each with
2 antennas. The advantages offered by Cl-Cl observations have been reviewed by Sasao \& Morimoto (1991) in the context of the Japanese VERA project.
As with conventional phase-referencing techniques, where antennas
are switched rapidly between a target and reference source, the
Cl-Cl mode is used to help elminate the unknown contributions
to the signal paths caused by the atmosphere and troposphere.
The neutral and ionized media lying between
a radio source and the surface of the Earth often have profound effects on
the radiation fields traversing them,
causing both temporal and spatial fluctuations in the signal path.
Such fluctuations depend strongly on the observing wavelength.
At cm and mm wavelengths, they are
predominantly caused by temporal fluctuations in the distribution of water
vapor in the troposphere.
The dominant causes at longer wavelengths,
however,
are irregularities, both small- and large-scale, in the electron density
distribution in the Earth's ionosphere. Most of the ionospheric effects scale
as $\nu^{-2}$. At approximately 2 GHz, the perturbations induced in the
incoming radiation as it propagates through the troposphere and ionosphere
have roughly equal magnitudes.

The standard phase-reference method using source-switching must make both
temporal and spatial interpolations between the target and
reference source phases. In contrast, in Cl-Cl mode observations
temporal variations above
each antenna, introduced in the wavefront towards each source, are continuously
followed, and no temporal interpolation is needed.
Thus, direction-dependent phase differences can be determined and corrected
for more accurately.
\section{Observations and correlation strategy}

We made 1.6 GHz Cl-Cl mode observations of 4 strong, compact radio sources,
selected from
the ICRF and VLBA calibrator lists.
The sources 
lie within a few degrees of each other on the sky (see Fig.~1).
We used 4 telescopes from the NRAO VLA (designated A, B, C, D), 4 separate 3-telescope
sub-clusters from the Westerbork (WSRT) array (W, X, Y, Z) and 3 telescopes from the Jodrell Bank MERLIN array
\begin{figure}[h]
 \centering
 \vspace{8.1cm}
 \includegraphics{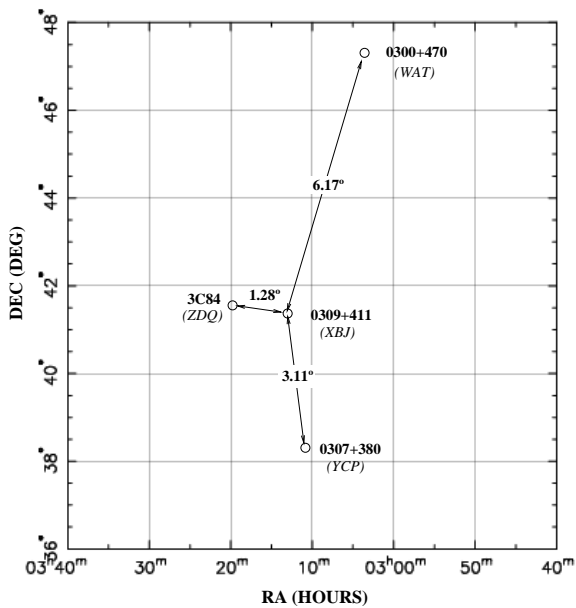}
  \caption{Positions of the 4 sources with their relative separations.
           In brackets, the triangle of antennas monitoring the source.}
   \label{rioja1}
\end{figure}
(J=Jodrell-Lovell, T=Tabley, P/Q=Jodrell-Mark2).
At each cluster site, each telescope tracked a different source continuously,
except for periods of about 7 minutes every 2 hours when all
telescopes observed the same calibrator source (3C84) for local calibration of the
arrays. In order to provide 4 separate sub-baseline triangles for
the 4 sources, the Mark2 telescope at Jodrell was switched every few minutes between 2 sources, to
compensate for the smaller number of elements in that cluster.
Thus, the 4 sources were observed with the 4 sub-triangles
WAT, XBJ, YCP, ZDQ.
The observations were made between UT 23:00 and 05:00, and
on 2 consecutive days (1999 November 16-18) in order to disentangle
tropospheric/ionospheric propagation effects from geometric errors.

In addition to the observations in Cl-Cl mode, simultaneous 
observations were made using a network of single telescopes
(Effelsberg, Medicina, VLBA-HN, NL, PT)
on the first day,
cycling between the 4 sources to permit mapping of 
their structures with a resolution similar to the observations in
Cl-Cl mode. This was designed to overcome the effect of poor $(u,v)$-coverage 
in the analysis of the Cl-Cl mode observations.

The observing and recording strategy were similar to those used in our
previous Cl-Cl observations at $\lambda$ 6 cm in June 1995 (Rioja et al.\ 
1997).
We recorded using MkIII Mode B, assigning different subsets of
video channels to different telescopes at each multi-element site.
Because the recorded bandwidth must be shared between the telescopes
in a cluster, we
recorded at double speed to improve the SNR on each sub-baseline.
At the VLA the 4 telescope signals were recorded using the 4 inputs of
a VLBA terminal, and a single MK4 terminal was used at WSRT (see Rioja, 1995
for the recording set-up).
For the MERLIN cluster the signals from Tabley and JB-Lovell were
recorded with a MK4 terminal, and the JB-Mark2 signal was recorded using a
VLBA terminal.
In this way we were able to record 4 x 4MHz channels for 3
of the sub-triangles, and 2 x 4MHz channels for triangle ZDQ.

The data were correlated using the MPIfR MkIIIa correlator in Bonn.
We followed special, non-standard procedures similar to those used in the previous 
Cl-Cl observations at $\lambda$ 6 cm,
using multiple passes to recover data from all the ``parallel'' sub-baselines
(WA, WT, AT, XB, XJ,..),
and from the ``cross''-baselines AX, AY, AZ, AJ and AP for the calibration
periods.
Amazingly, given the complexity of the observing, recording and
correlation set-up, fringes were indeed detected on all sub-baselines !
\section{Preliminary post-processing analysis}

The post-correlation data reduction is being performed using the AIPS package.
A number of contributions to the signal path were missing in the {\it a priori}
model used in the MkIIIa correlator and we used the task CLCOR to
correct for the telescope axis offsets and parallactic 
angle contributions to the observed phases, delays and rates.
We also used CLCOR to implement improved station positions for WSRT (tied-array position)
and JB-Mark2 in ITRF2000, derived from a special $\lambda$ 6 cm experiment made using geodetic
techniques (Charlot et al. 2001). These have estimated accuracies better than 10 cm.
We used locally determined telescope separations to derive IRTF2000 positions
for JB-Lovell and Tabley.
However, the biggest uncorrected contribution in observations at 1.6 GHz
is due to the propagation of the wavefronts through the ionosphere.
This is particularly large during a maximum in the cycle of solar activity.
In fact the epoch of these observations was very close to the last solar
maximum.

The preliminary analysis has followed two paths:
\subsection{Standard analysis}

We have made a separate phase self-calibration analysis
for each of the 4 sources, observed on the 4 sub-baseline triangles.
From the individual analyses of observations of the 4 triangles,
we find that the delay and rate residuals of telescopes belonging to the same
``cluster'' do show a similar behaviour.
\begin{figure}[h]
 \centering
 \vspace{8cm}
 \includegraphics{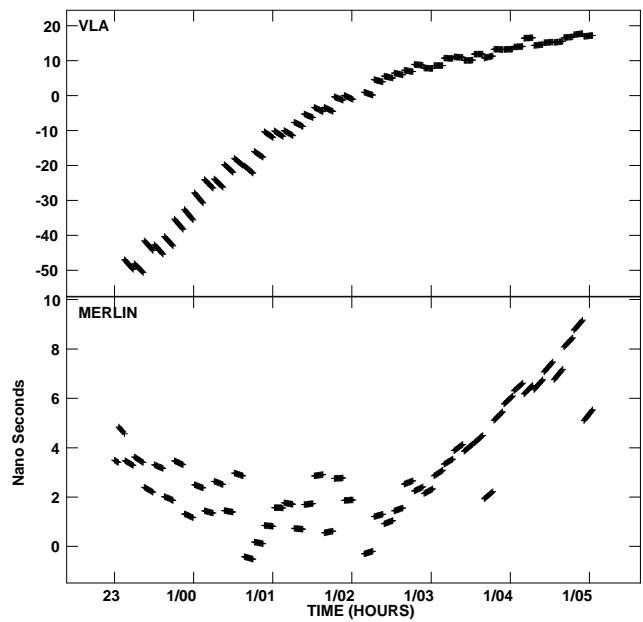}
  \caption{Instrumental delays estimated with FRING for telescopes from
   the VLA and MERLIN clusters, using WSRT as reference. The VLA residuals
   have the signature of a dispersive medium, like the ionosphere, until
   about 03:00 UT. See text for further explanation.}
   \label{rioja2}
\end{figure}

Fig.~2 shows an example of the temporal behaviour of the
residual delay (determined using FRING) at the VLA and MERLIN
sites, using the WSRT site as a reference (and before the
antenna position corrections were made).
It is illustrative to note the opposite signs of the delay rate ---determined
from the phase rate assuming a non-dispersive medium--- (the individual slopes
within each dashed segment) and the ``true'' delay rate (slope between consecutive
dashed segments) for the VLA site, from the beginning of the
observations at 23:30 UT until about 03:00 UT.  This behaviour is
characteristic of a dispersive medium, where the group delay
produces a phase change of opposite sign compared with that from the same delay
produced by a geometric or neutral atmospheric path error. We conclude
that during this period the residual delay must be largely caused by the ionosphere.
However, for the MERLIN site, the phase and delay rate behaviour with
time agree, showing that the ionosphere is not the main source of
delay error on the short MERLIN-WSRT sub-baselines, at least at night time.
(Note that, in this plot, there are still residual geometric errors due to poor
antenna coordinates).  Note also the coincidence between the Sun setting at
the VLA, and the agreement between the phase and
delay rates after ca. 03:00 UT ---clearly the ionosphere has shut down !
\subsection{Phase reference analysis}

We have also attempted a preliminary phase referencing analysis.
Analysis of conventional ``nodding'' phase reference observations
can require a
double interpolation of the instrumental residual phases, in time (between
successive calibrator observations) and perhaps also in the sky (between calibrator
positions if attempting a spatial interpolation ---see Fomalont \& Kopeikin, {these Proceedings})
to estimate the phase correction at the location of the target
source.
In contrast, phase-referencing in Cl-Cl observing mode consists of a spatial
interpolation of the instrumental residual phases between simultaneous
observations of calibrators in different directions.

We selected a ``reference triangle'' (XBJ) and its source (0309+411, the ``reference source'')
and transferred their phase solutions, using phase-reference
techniques, to the analysis of the observations from the other 3 triangles (``target
triangles'') and their ``target sources''.
This strategy is similar to that used for the in-beam phase-referencing used
for the analysis of the close source pair 1038+528\,A,B (Rioja and Porcas, 2000).
Even if there are no geometric errors arising from poor antenna and/or source coordinates,
the spatial structure of the ionosphere
will affect the referenced phases.
Figure 3 shows the referenced visibility phases for
the 3 sub-baselines,
for each of the 3 target triangles observing a
target source.
As expected, there is a strong correlation between
the rate of change of the referenced visibility phases with time
on the various sub-baselines, and the increasing angular separation between
target and reference source (1\fdg28 for 3C\,84,
3\fdg11 for 0307+380 and 6\fdg17 for 0300+470).
This is, as expected, particularly noticeable
on the longest baselines to the VLA during the first half of the
observations.
\section{Summary and further work} 

The complex observations and processing of the Cl-Cl mode VLBI observations
between 3 ``cluster'' sites, presented in this contribution, have
been successful. The resulting data set provides an excellent
opportunity to investigate the capabilities of this new
technique at low frequencies,
and to explore new analysis routes.
The potential scientific returns may be realised in the near future
when new, dedicated
instruments such as VERA come into operation.
\begin{figure}[t]
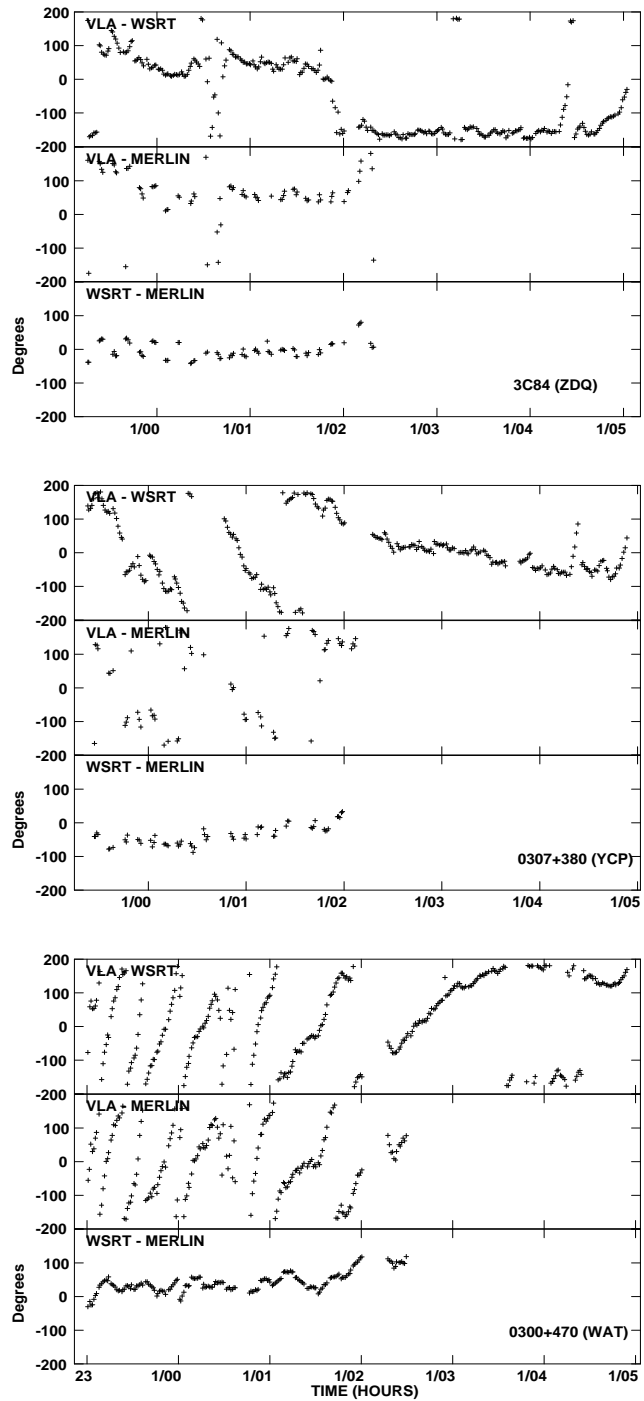

 \centering
 \vspace{6.5cm}
 \includegraphics{MRioja_fig3.ps}
 \centering
 \vspace{6.3cm}
 \includegraphics{MRioja_fig4.ps}
 \centering
 \vspace{6.3cm}
 \includegraphics{MRioja_fig5.ps}
  \caption{Referenced visibility phases for the observations 
    of target sources with target triangles, as indicated
          in each plot.
    In the phase-reference analysis
    we used observations of 0309+411 on the triangle XBJ as reference.
    The change of phase with time scales with source/reference angular 
    separation (from top to bottom, 1\fdg28, 3\fdg11 and 6\fdg17)}
   \label{rioja3c}
\end{figure}

At present, we are continuing to work for one of the main goals of these observations
in Cl-Cl mode, to determine the size of isoplanatic regions in the ionosphere
and the effects on VLBI observations.
With a 3 calibrator experiment (or 2 calibrators if
they lie on a straight line with the target source), one can remove
both temporal and spatial phase fluctuations.  The problem with conventional ``nodding''
between 3 or 4 sources is that the temporal phase interpolation between sources is very
difficult because of the large gaps between observations of a given source.
It is hoped that the Cl-Cl method can provide accurate phase measurements
by observing 3 calibrator sources simultaneously and removing a
2D gradient with angle on the sky from the data of the 4th (target) source.
This would remove the effect of the relatively large scale ionospheric
refraction on
the sky (isoplanatic ionospheric effects), particularly at low frequencies.
\begin{acknowledgements}
We would like to thank all the people involved in the preparations
for this {\it highly} non-standard observation at each site, and the
staff at the MPIfR correlator. 
Their altruist efforts and great expertise have always played a major role
in our series of VLBI observations in Cluster-Cluster mode - 
to the point that none of this would be possible otherwise.  
The European VLBI Network is a joint facility of European, Chinese
and other radio astronomy institutes funded by their national research councils.
NRAO is a facility of the US National Science Foundation, operated by
Associated Univerties Inc.
M.J.R., L.I.G.\ and R.T.S.
acknowledge partial support from the EC ICN
RadioNET (Contract No. HPRI-CT-1999-40003).
\end{acknowledgements}

\end{document}